\documentclass{PoS}

\usepackage{amsmath,amsfonts,amssymb}
\usepackage{mathtools}
\usepackage{braket}
\usepackage{multirow}
\newcommand{\be}{\begin{equation}}
\newcommand{\ee}{\end{equation}}

\def\cO{{\cal O}}

\title{Black holes in Asymptotically Safe Gravity}

\ShortTitle{Black holes in Asymptotically Safe Gravity}

\author{\speaker{Frank Saueressig}, Natalia Alkofer, Giulio D'Odorico, Francesca Vidotto%
         \thanks{The work by F.S., G.D.\ and N.A.\ is supported by the FOM-grant 13PR3137.  F.V.\ acknowledges support from the Netherlands
Organisation for Scientifc Research (NWO) Veni Fellowship Program}\\
        Institute for Mathematics, Astrophysics and Particle Physics (IMAPP) \\
        Radboud University Nijmegen, The Netherlands\\
        E-mail: \email{f.saueressig@science.ru.nl} \\
E-mail:  \email{n.alkofer@science.ru.nl} \\
E-mail: \email{g.dodorico@science.ru.nl} \\
E-mail: \email{f.vidotto@science.ru.nl} \\}

\abstract{
Black holes are among the most fascinating objects populating our universe. Their characteristic features, encompassing spacetime singularities, event horizons, and black hole thermodynamics, provide a rich testing ground for quantum gravity ideas. In this note we observe that the renormalization group improved Schwarzschild black holes constructed by Bonanno and Reuter within Weinberg's asymptotic safety program constitute a prototypical example of a Hayward geometry used to model non-singular black holes within quantum gravity phenomenology. Moreover, they share many features of a Planck star: their effective geometry naturally incorporates the one-loop corrections found in the effective field theory framework, their Kretschmann scalar is bounded, and the black hole singularity is replaced by a regular de Sitter patch. The role of the cosmological constant in the renormalization group improvement process is briefly discussed.
}

\FullConference{Frontiers of Fundamental Physics 14\\
                 15-18 July 2014\\
                 Aix Marseille University (AMU) Saint-Charles Campus, Marseille, France}

\begin{document}
%--------------------------------------------------------------------------
%\section{Introduction}
%--------------------------------------------------------------------------
Black holes have now become objects routinely observed in astrophysics
\cite{Falcke:2013ola}.  General relativity describes very well their exterior, as well as
their horizon and interior, up to the central singularity, where the theory
fails.  On physical grounds, we expect the physics of the deep central
region to be strongly affected by quantum effects, therefore using general
relativity all the way up to the singularity is pushing the theory outside
its domain of validity. A theory reconciling general relativity with
quantum mechanics is needed to describe this central region.
A rather conservative proposal to embed gravity in the quantum field theory framework 
is Weinberg's Asymptotic Safety scenario \cite{Weinberg:1980gg}. The key ingredient 
for investigating this scenario is the gravitational effective average action $\Gamma_k$ \cite{Reuter:1996cp}.
By construction, $\Gamma_k$ is a Wilson-type action functional whose effective vertices already incorporate quantum fluctuations with momentum $p^2 \gg k^2$. Thus $\Gamma_k$ can be thought of as providing an effective description of gravitational phenomena on typical momentum scales $k$.
As its central property, the effective average action satisfies a formally exact functional renormalization group equation,
which, by now, has accumulated substantial evidence that the gravitational renormalization group (RG) flow possesses
a non-trivial fixed point (NGFP) which could provide the UV completion
of gravity at trans-Planckian energies, see \cite{Percacci:2011fr,Reuter:2012id} for reviews.

\begin{figure}[t]
\begin{center}
  \includegraphics[width=0.75\textwidth]{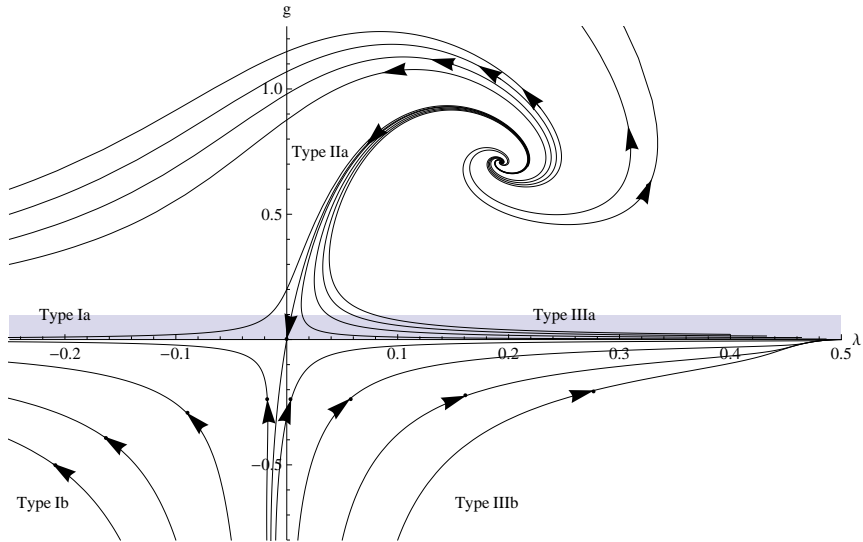}
\end{center}
\caption{\label{phasedia}
Phase diagram showing the gravitational RG flow of the Einstein-Hilbert truncation in terms of the dimensionless coupling constants $g_k \equiv G_k k^{2}$ and $\lambda_k \equiv \Lambda_k k^{-2}$. The flow is governed by the interplay of the NGFP located at $g_* > 0, \lambda_* > 0$ and the GFP at the origin. The arrows point towards the infrared, i.e., in the direction of lower $k$-values. Adapted from \cite{Reuter:2001ag}.}
\end{figure}

The RG flow obtained by approximating $\Gamma_k$ by the Einstein-Hilbert action
\be
\Gamma_k^{\rm grav} = \frac{1}{16 \pi G_k} \int d^4x \sqrt{g} \left( 2 \Lambda_k - R \right) 
\ee
with scale-dependent (dimensionless) Newton's constant $g_k \equiv k^2 \, G_k$ and cosmological constant $\lambda_k \equiv \Lambda_k/k^2$
is shown in Fig.\ \ref{phasedia}. The phase diagram depicts the NGFP at positive $g_* > 0, \lambda_* > 0$ which acts as the UV completion of all RG trajectories with positive Newton's constant. Lowering the RG scale $k$ the flow undergoes a crossover towards the Gaussian fixed point (GFP) situated in the origin. In the vicinity of the GFP the {\it dimensionful} coupling constants $G_k$ and $\Lambda_k$ become independent of $k$, so that classical general relativity is recovered in the IR. Depending on whether the RG trajectory ends at the GFP or flows to its left (right) a zero or negative (positive) IR value of the cosmological constant is recovered. The scaling of the coupling constants at the NGFP is easily deduced from the dimensionless couplings becoming constant,
\be\label{NGFPscaling}
\mbox{NGFP:} \qquad G_k = g_* \, k^{-2} \, , \quad \Lambda_k = \lambda_* \, k^2 \, , 
\ee
while in the IR close to the GFP \cite{Reuter:1996cp}
\be\label{lowenergy}
G_k = G_0 \left( 1 - \omega \, G_0 \, k^2 + \cO(G_0^2 \, k^4) \right) \, ,
\ee
with $\omega > 0$ a fixed number dependent on the particular choice of regularization scheme.
For the regulator used in Fig.\ \ref{phasedia}, $\omega = 11/6\pi$.
%

%--------------------------------------------------------------------------
%\section{Renormalization group improving classical black hole solutions}
%--------------------------------------------------------------------------
Classical Schwarzschild black holes are exact vacuum solutions of Einstein's
field equations. The geometry is characterized by the line element
\be\label{ssmetric}
ds^2 = -f(r) \, dt^2+ f(r)^{-1} \, dr^2 + r^2 d\Omega_2^2
\ee
with $d\Omega_2^2$ denoting the line-element of the two-sphere and the radial function
\be\label{frfct}
f(r) = 1 - \frac{2 \, G \, m}{r} \, . 
\ee 

Following \cite{Bonanno:1998ye,Bonanno:2000ep} quantum gravity corrections to the classical black hole geometry may be incorporated by RG improving the classical solution.\footnote{For an investigation of the RG improved black holes from the perspective 
of black hole thermodynamics see \cite{Falls:2012nd}. A recent review with
further references can be found in \cite{Koch:2014cqa}.}
The basic idea underlying the RG improvement is to promote the constant $G$ to depend on the renormalization group scale $k$, replacing $G \mapsto G_k$. Subsequently, the RG
group scale is identified with a physical scale of the (classical) geometry. For the spherically symmetric Schwarzschild solution, it is natural to relate $k$ to 
 the absolute value of the radial proper distance $d_r(r)$ between a point $P(r)$ in the spacetime and the center of the black hole
\be
d_r(r) = \int \sqrt{|ds^2|} \, .
\ee
Close to the origin and at asymptotic infinity, $d_r(r)$ has the expansions
\be\label{asympkvonP}
d_{r}(r)|_{r\ll 2 \, G_0 \, m} \simeq \frac{2}{3}\frac{1}{\sqrt{2 G_0 m}} \, r^{3/2} + {\mathcal{O}}(r^{5/2}) 
\, , \qquad
d_{r}(r)|_{r\gg 2 \, G_0 \, m} \simeq r+ {\mathcal{O}}(r^{0}) \, .  
\ee
The cutoff identification then relates the momentum scale $k$ to this distance according to\footnote{There is no predetermined recipe for determining the identification of $k$ with a physical scale of the system. An equally valid choice relates $k$ to the proper time measured by a freely falling observer starting at $P(r)$ to reach the black hole singularity. Notably, these alternative choice
 lead to similar results as the ones reported below.}
\be\label{kvonP}
k(r)=\frac{\xi}{d_r(r)}\, , 
\ee
with $\xi$ being a free parameter. Applying this RG improvement procedure to the classical radial function \eqref{frfct}
yields the RG improved geometry where $f(r)$ is given by
\be\label{frimp}
f(r) = 1 - \frac{2 \, G(k(r)) \, m}{r} \, . 
\ee
Thus the RG improvement procedure promotes the classical Schwarzschild metric
to a Hayward-type effective geometry \cite{Hayward:2005gi} with $f(r) = 1 - 2 M(r)/r$, where the function
$M(r)$ is determined from an RG trajectory constructed within the fundamental theory 
and the RG improvement \eqref{kvonP}.

\begin{figure}[t]
\begin{center}
  \includegraphics[width=0.48\textwidth]{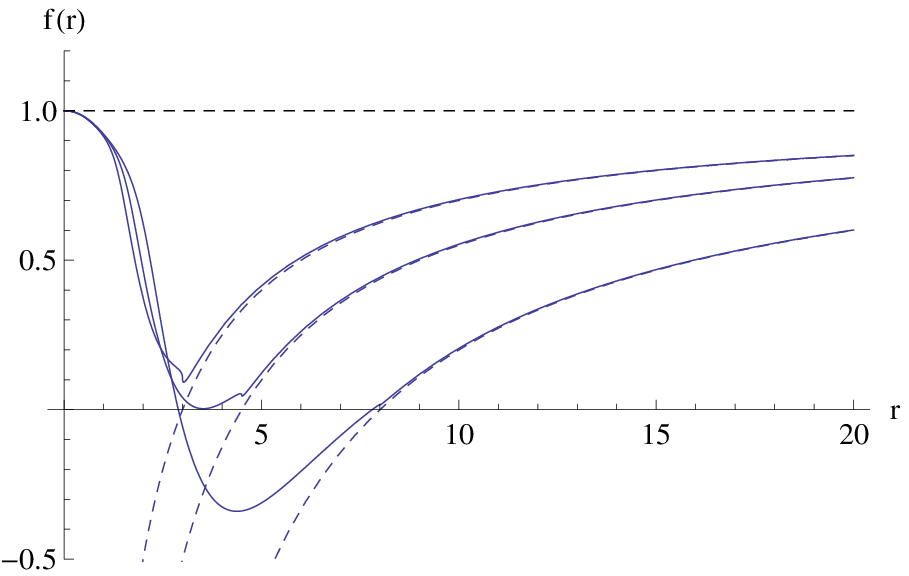} \quad
  \includegraphics[width=0.48\textwidth]{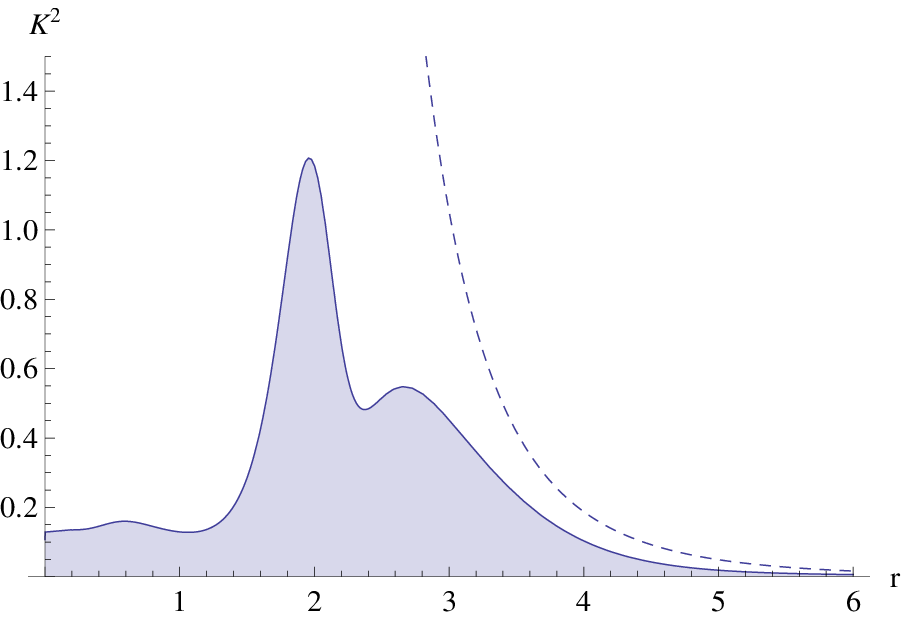}
\end{center}
\caption{\label{p.horizon}
The left diagram illustrates the horizon structure for the RG improved Schwarzschild black holes with $m = 1.5$ (top curve), $m = m_{\rm crit} \approx 2.25$ (middle curve) and $m = 4$ (bottom curve), while the Kretschmann scalar curvature $K^2 = R_{\mu\nu\rho\sigma} R^{\mu\nu\rho\sigma}$ for the case $m = 4 > m_{\rm crit}$ is shown in the right diagram. All quantities are measured in Planck units. The classical result is visualized by the dashed curves for comparison.}
\end{figure}
In the asymptotic regimes of the black hole spacetime,
the effect of the RG improvement can be traced analytically. Substituting
the low energy expansion \eqref{lowenergy} into \eqref{frimp} and evaluating the cutoff
identification \eqref{kvonP} for small $k$, yielding $k^2 = \xi^2/r^2 + \cO(r^{-3})$,
results in
\be
\left. f(r) \right|_{r\gg 2 \, G_0 \, m} \simeq 1 - \frac{2 G_0 m}{r} \left( 1 -  \frac{\tilde{\omega} \, G_0}{r^2} \right) \, , 
\ee 
with $\tilde \omega = \omega \xi^2$. The improved line-element naturally incorporates 
the 1-loop corrections found in effective field theory \cite{Donoghue:1993eb} and can be matched
by adjusting the free parameter in the cutoff identification to be $\xi^2 = \tilde \omega/\omega$.
Using $\omega = 11/6\pi$ and $\tilde \omega = 118/15\pi$ we obtain $\xi^{\rm 1-loop} \approx 2.07$,
which we will use in numeric evaluations below. Close to the black hole singularity, the RG improvement is based on the fixed point scaling \eqref{NGFPscaling}.
Substituting the asymptotic cutoff identification based on \eqref{asympkvonP} then yields
\be
\left. f(r) \right|_{r\ll 2 \, G_0 \, m} \simeq 1 - \tfrac{1}{3} \, \Lambda_{\rm eff} \, r^2 \, , \quad \mbox{with} \qquad \Lambda_{\rm eff} = \frac{4}{3} \frac{g_*}{G_0 \xi^2} \, . 
\ee 
Thus the RG improvement correctly incorporates the one-loop corrections
determined in effective field theory (fixing the only free parameter in the procedure)
and resolves the black hole singularity by giving rise to a de Sitter type behavior close to the center.

The complete RG improved radial function can easily be constructed numerically. For concreteness we choose the underlying RG trajectory to be the Type IIa trajectory (see Fig.\ \ref{phasedia}) connecting the GFP with the NGFP, setting $\Lambda_0 = 0, G_0 = 1$. The resulting improved $f(r)$ depends on the asymptotic mass of the black hole $m$ only and is shown in the left diagram of Fig.\ \ref{p.horizon}. For $m > m_{\rm crit}$ the improved geometry has an outer and inner horizon. For $m = m_{\rm crit}$, with $m_{\rm crit}$ being of the order to the Planck mass, the two horizons coincide while for $m < m_{\rm crit}$ there is no horizon. The Kretschmann scalar curvature of the improved geometry (right diagram) peaks below the inner horizon and its maximum value is (approximately) given by the Planck scale.

\begin{figure}[t]
\begin{center}
  \includegraphics[width=0.48\textwidth]{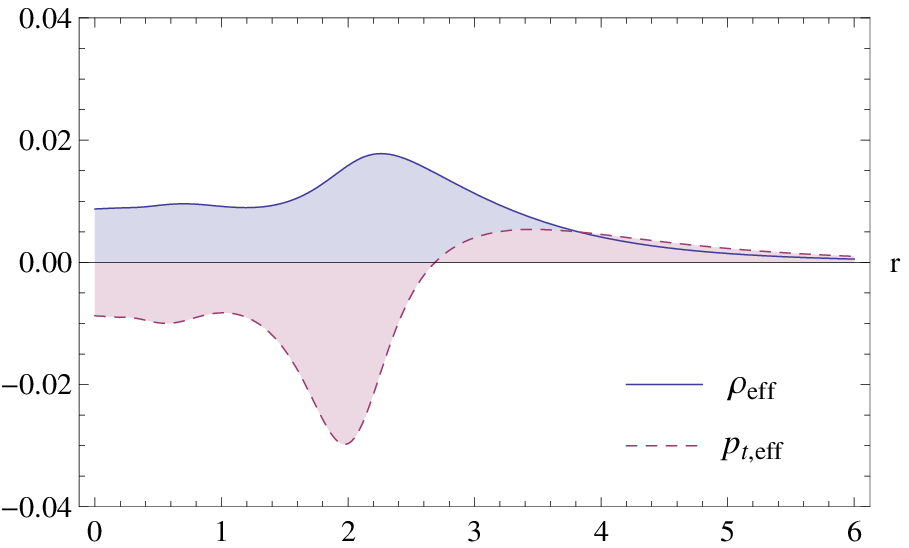} \quad
  \includegraphics[width=0.48\textwidth]{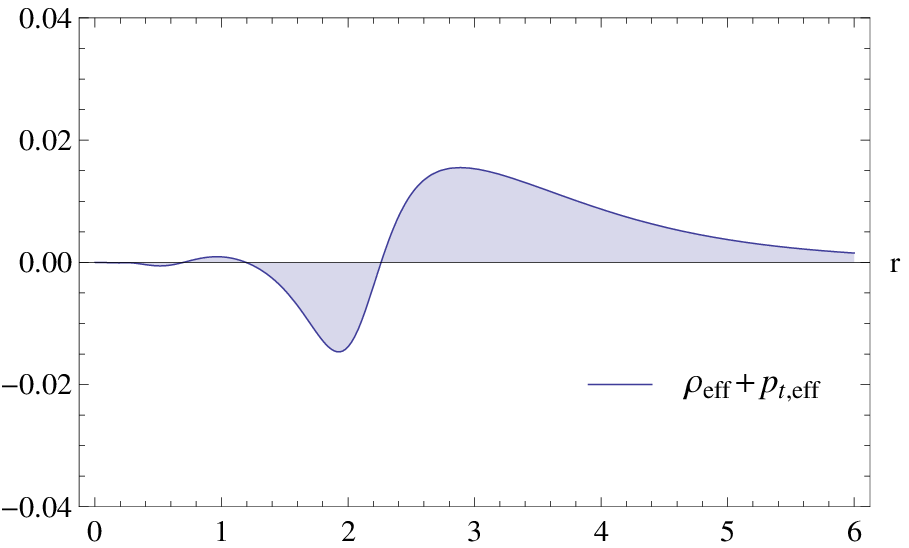}
\end{center}
\caption{\label{p.wec}
Effective energy density and pressure profiles for the RG improved Schwarzschild black hole with $m=4, G_0 = 1$. The radial component of the weak energy condition is zero everywhere, while the transversal contribution, shown in the right diagram, violates the condition on scales below the inner horizon.
}
\end{figure}
Substituting the RG improved geometry into the {\it classical} Einstein equations allows to interpret the resulting modifications in the classical black hole geometry as a quantum contribution to the energy momentum tensor. The resulting effective energy density $\rho_{\rm eff}$ and transverse pressure $p_{t, {\rm eff}}$ are shown in the left diagram of Fig.\ \ref{p.wec}. The radial pressure $p_{r, {\rm eff}} = - \rho_{\rm eff}$, so that the RG improvement acts like a cosmological constant in the radial direction. The right diagram of Fig.\ \ref{p.wec} displays the weak energy condition $\rho_{\rm eff} + p_{\rm eff}$ for the effective energy momentum tensor. Notably, the weak energy condition is violated at subhorizon scales due to strong transversal pressure.

Recently, Ref.\ \cite{Koch:2013owa} generalized the study of the RG improved Schwarzschild black hole to Schwarzschild-(Anti-) de Sitter
black holes, with the radial function \eqref{frfct} also including a non-zero cosmological constant
\be\label{frcosmo}
f(r) = 1 - \frac{2 \, G \, m}{r} - \frac{1}{3} \, \Lambda \, r^2 \, . 
\ee
This extension is motivated by the observation that, even for the case where $\Lambda_0 = 0$
a non-zero cosmological constant will be generated along the RG flow (cf.\ Fig.\ \ref{phasedia}).
Moreover, in the vicinity of the NGFP the scaling \eqref{NGFPscaling} implies
that the {\it dimensionful} Newton's constant goes to zero while the dimensionful cosmological constant
actually diverges when $k \rightarrow \infty$. Thus, contradicting the intuition that the cosmological 
constant is important at large distances only, its inclusion may also influence the structure
of microscopic black holes. Indeed, applying the RG improvement procedure for the
Schwarzschild case to the radial function \eqref{frcosmo} and evaluating the result 
for the fixed point scaling \eqref{NGFPscaling}  
the RG improved line-element \emph{valid at the NGFP} is again of the form \eqref{frcosmo}
\be
f_*(r) = 1  - \frac{2 \, G_0 \, m}{r} \left( \frac{3}{4} \lambda_* \xi^2  \right) - \frac{1}{3} \, \left( \frac{4 g_*}{3 G_0 \xi^2} \right) \, r^2 \, .
\ee
Thus the RG improved Schwarzschild-de Sitter black holes become self-similar in the sense that 
the line-element takes the same form in the IR and UV.
Notably, the inclusion of the scale-dependent cosmological constant has also reintroduced a singularity which, for the case of the Schwarzschild black hole, has been removed by the RG improvement process. At this stage it is worth stressing that the physical nature of this new singularity is actually quite different to the one found in the classical black hole solution: applying the RG improvement procedure to flat Lorentzian spacetime also introduces a singular behavior of the RG improved line-element even in the absence of any matter. This nurtures the speculation that the ``quantum'' singularity introduced by the cosmological constant may actually reflect a feature of quantum spacetime which is actually unrelated to the study of black holes \cite{Koch:2014cqa}.

%--------------------------------------------------------------------------
%\section{Conclusions and outlook}
%--------------------------------------------------------------------------

In summary, the RG improved Schwarzschild black holes found within
Asymptotic Safety \cite{Bonanno:1998ye,Bonanno:2000ep} naturally fall into the class of Hayward metrics \cite{Hayward:2005gi}
which have been proposed as effective models for non-singular black holes.
The disappearance of the central singularity has also been observed
in other approaches to quantum gravity, as, e.g., in Loop Quantum
Gravity.   This physical scenario has been recently investigated in 
\cite{Rovelli:2014cta,DeLorenzo:2014pta,Haggard:2014rza}, where the non-singular central core of the black hole is called
``Planck star''. Interestingly, this opens a new window for quantum gravity
phenomenology \cite{Barrau:2014hda,Barrau:2014yka} as the disruption of the horizon should give a
characteristic astrophysical signal. We hope that this link between
fundamental theories of gravity and effective geometries will be useful
towards improving our understanding of black-hole evolution.

\end{document}